\documentclass[10pt,leqno]{amsart}
\usepackage[utf8]{inputenc}
\usepackage[T1]{fontenc}
\usepackage{graphicx}
\usepackage{amsmath,amssymb,amsthm}
\usepackage{booktabs}
\usepackage{array}
\usepackage{url}
\usepackage{xcolor}

\graphicspath{{figures/}}

\newtheorem{theorem}{Theorem}
\newtheorem{lemma}{Lemma}
\newtheorem{proposition}{Proposition}

\theoremstyle{definition}
\newtheorem{definition}{Definition}
\newtheorem{remark}{Remark}
\newcommand{\R}{\mathbb{R}}
\newcommand{\sat}{\operatorname{sat}}
\newcommand{\col}{\operatorname{col}}

\title[Privileged Control to Deployable Adaptation]{From Privileged Control to Deployable Adaptation:\\Fusing Mechanism-Guided Task Reduction with Learned Behavior}
\author{Xitong Niu}
\author{Peifeng Hui}
\author{Zheyong Jiang}
\author{Yuan Gao}
\author{Chuanlin Zhang}
\address{Shanghai University of Electric Power, Shanghai, China}
\email{xitongniu@mail.shiep.edu.cn}
\date{August 2026}

\begin{document}
\begin{abstract}
Simultaneous input-gain variation and large additive disturbance create a
control problem in which a fixed observer or nominal controller may be unable
to reproduce the performance of a regime-aware design.  We study a
training--deployment asymmetry: during simulation or commissioning, an expert
controller is allowed to use the known gain and disturbance, whereas the
deployed controller can use only the reference and measured states.  Directly
imitating expert actions is generally unsafe because the same instantaneous
student observation may correspond to different privileged regimes and hence
different expert actions.  We propose a mechanism-guided transfer route rather
than a new neural architecture.  An exact sampled-data identity removes the
additive disturbance from the expert law and reduces learning to a
task-relevant inverse input gain inferred from causal state history.  The
latent target is reconstructed from expert actions and deployment-visible
trajectories, so the true plant parameter is not required as a student label.
A common-quadratic certificate is derived for the actual augmented sampled
recursion, followed by explicit residual, coverage, switching, noise, and
saturation qualifications.  A parameter-regime scan shows that the nominal
observer's error grows sharply as $a$ decreases and that the next gain above the
best non-failing tuning diverges for every tested $a<1$.  Direct action networks
also fail in closed loop despite moderate offline error, whereas the structured
student remains close to the privileged expert and reduces tracking RMSE by
about 69\% relative to the tuned observer in unseen 60-s trials.  The
contribution is an interpretable design perspective for
turning privileged multi-regime control knowledge into a deployable adaptive
controller, together with conditions under which the transfer is meaningful.
\end{abstract}
\maketitle
\enlargethispage{4pt}

\section{Introduction}

Unknown disturbances and plant-parameter changes often appear together, but
they do not enter a plant in the same way.  An additive disturbance translates
a state derivative, while a multiplicative uncertainty changes how strongly a
state or control channel acts.  A controller tuned for one nominal model must
therefore trade tracking, gain robustness, disturbance rejection, measurement
noise, and actuator demand.  Disturbance-observer-based control (DOBC) is a
powerful and mature response to uncertainty \cite{chen2016dobc}; nevertheless,
estimating one lumped disturbance with a fixed bandwidth is not equivalent to
having the changing input gain and disturbance separately available.

This paper changes the information pattern during controller development.  In
simulation, commissioning, or product validation, operating conditions can be
deliberately imposed and recorded.  We allow a high-performance controller to
use these known conditions and call it a \emph{privileged expert}.  Its purpose
is to generate a family of closed-loop demonstrations, not to be deployed.
The deployed student sees only ordinary measurements and the reference.  The
central question is:
\begin{quote}
Can privileged multi-regime control knowledge be converted into a single
deployable controller without giving privileged variables to that controller?
\end{quote}

The apparently simple answer---regress the expert action on measured
states---contains a structural trap.  If two hidden regimes produce nearly the same student observation but require different expert actions, no deterministic instantaneous student can reproduce both.  A larger network cannot restore missing information.  Moreover, low one-step regression error need not imply a good feedback controller because the learned action changes the future input distribution, the classical covariate-shift issue in imitation learning \cite{ross2011dagger}.  These two facts explain why we begin with mapping realizability and closed-loop structure rather than architecture selection.

For the second-order system studied here, causal history contains a signature of the hidden regime.  More importantly, the plant and expert equations yield an exact sampled identity: the additive disturbance can be eliminated algebraically, leaving only a task-relevant inverse input gain to learn.  A small feedforward network estimates this latent factor from a sparse history; a fixed physics layer reconstructs the action.  Thus the neural network behaves as a task-oriented online identifier, but it is trained from privileged expert behavior rather than from a prescribed observer trajectory.  The original large-signal gain--disturbance problem becomes a known feedback law plus a bounded latent-estimation error.

\begin{figure}[!t]
\centering
\includegraphics[width=\textwidth]{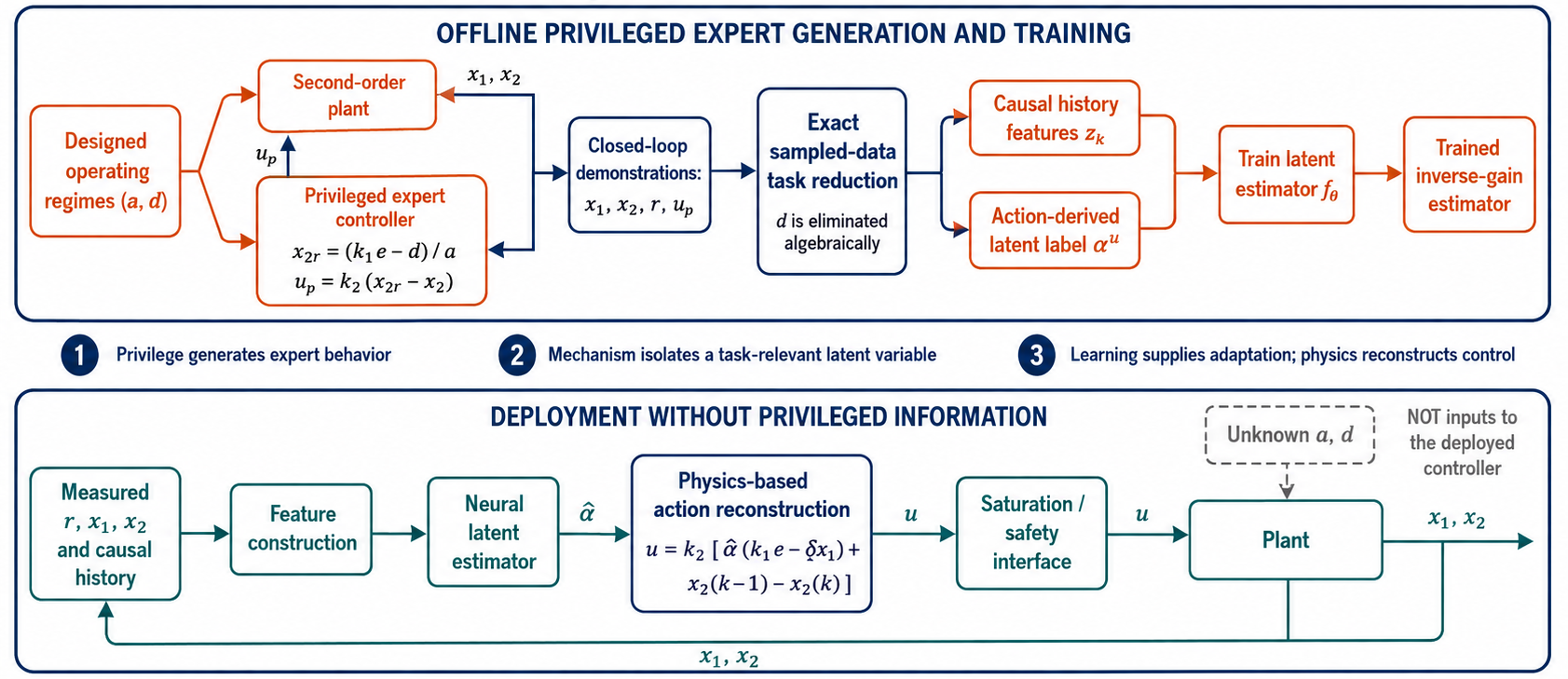}
\caption{Overall information flow.}
\label{fig:framework}
\end{figure}

Figure~\ref{fig:framework} summarizes the resulting information architecture:
privileged variables appear only in offline expert generation, whereas the
deployed path contains measurable history, the learned inverse-gain latent,
and the fixed physical action reconstruction.

The paper's contributions are:
\begin{enumerate}
\item an explicit privileged-training/deployable-inference formulation for
simultaneous multiplicative and additive uncertainty;
\item a realizability diagnosis showing why direct instantaneous imitation can
be ill posed and why histories must be justified by information, not merely
added as network capacity;
\item an exact sampled-data task reduction and an action-derived latent label
that does not use the true plant parameter as student supervision;
\item a step-by-step common-Lyapunov analysis of the actual delayed sampled
recursion, including bounded implementation residuals and transparent limits;
\item ablations and multi-regime comparisons against a privileged expert and a
gain-swept disturbance observer, including unseen long-horizon trials.
\end{enumerate}
The specific 64--32 neural network is intentionally ordinary.  The claimed
innovation lies in the information design, task reduction, training route,
interpretability, and conditional verification.  The analysis also states
what is \emph{not} proved, so that the second-order example provides a precise
starting point for higher-order and experimental extensions.

The distinctive point is the conjunction of five elements rather than any one
ingredient in isolation: privileged variables are used to construct a
closed-loop expert distribution; those variables are prohibited at student
deployment; a neural estimator uses causal history to infer only a
task-relevant latent factor; an exact physical layer reconstructs the action;
and the resulting sampled loop is analyzed through an explicit relative-gain
certificate.  Privileged policy learning, neural adaptive control,
physics-guided learning, and learned-control verification each have substantial
literatures.  Our claim is that their composition through an
\emph{action-derived, physically factorized latent target} addresses a control
information mismatch that none of those ingredients resolves alone.  The
comparison in Section 2 states this positioning explicitly.

\paragraph{Organization.}
Section 2 positions the work.  Section 3 defines the information pattern and
realizability problem.  Sections 4--5 derive the structured student and its
training labels.  Section 6 gives the stability analysis, Section 7 reports
simulations, and Sections 8--9 discuss limitations and supply detailed
derivations and reproducibility information.

\section{Related Work and Positioning}

\subsection{Learning with privileged information}
Learning using privileged information (LUPI) formalizes the idea that training
may access variables unavailable at inference \cite{vapnik2009lupi}.  In
robotics, ``learning by cheating'' uses privileged simulator representations to
train a deployable vision policy \cite{chen2020learning}.  Rapid Motor
Adaptation separates a base policy from a history-based adaptation module for
unmeasured environmental properties \cite{kumar2021rma}, and related
whole-body control work uses latent adaptation for complex locomotion and
manipulation \cite{fu2023deep}.  A-NC trains a recurrent deployable controller
from trajectories generated by a parameter-aware nonlinear model predictive
controller \cite{paluch2025anc}.  These studies establish the broad value of
privileged training; our narrower contribution is an algebraically structured
control transfer with a numerical sampled-data certificate.

Recent work also questions whether privileged variables are always the best
route.  SLR learns quadruped locomotion without privileged information
\cite{chen2025slr}, emphasizing the cost of hand-selected or explicitly
estimated privileged states.  This does not contradict our setting: the
operating variables are available during designed simulation/commissioning
tests, and we use them to construct the expert but remove them from both the
recommended student label and deployment interface.

Privileged teachers can be unrealizable for students with weaker observations.
Teacher Guided Reinforcement Learning explicitly separates what a teacher can
do from what the student can represent \cite{shenfeld2023tgrl}.  Recent theory
also characterizes conditions under which expert distillation can succeed in
partially observed systems \cite{cai2024provable}, while realizable-student
distillation studies recovery from full-state teachers that prescribe actions
outside a partially observed student's information class
\cite{kim2025realizable}.  Our state--action conflict test is the deterministic
control analogue of this information asymmetry.

\subsection{Imitation and distribution shift}
Policy distillation compresses one or more policies into a student
\cite{rusu2015policy}.  Behavioral cloning, however, is trained on the
expert-induced state distribution.  DAgger shows why sequential prediction
must account for the distribution induced by the learned policy
\cite{ross2011dagger}; a broader algorithmic treatment is given in
\cite{osa2018imitation}.  We do not run interactive DAgger because the present
goal is an offline, reproducible transfer from existing Simulink trajectories.
Instead, we use closed-loop ablations and long-horizon rollouts to expose the
gap between one-step imitation and control performance.

\subsection{Observers, data-driven control, and learned certificates}
DOBC and related methods estimate and compensate uncertainty within an
explicit feedback design \cite{chen2016dobc}.  Model-free adaptive control
\cite{hou2013mfac} and behavioral/data-driven formulas
\cite{depersis2020formulas} represent other ways to extract control-relevant
information from data.  Our method is not model free: it deliberately retains
the known relative-degree structure and learns only the factor that the
mechanism cannot supply at deployment.  This is closer to mechanism-guided
learning than to a black-box replacement of control theory.

Learned controllers can be paired with Lyapunov or barrier certificates, but
the certificate must match the implemented dynamics and its assumptions
\cite{dawson2023survey}.  Here the delayed backward difference creates a
three-state augmented recursion.  We derive that recursion exactly and verify a
common quadratic inequality at the endpoints of the learned relative-gain
interval.  The result is regional and excludes saturation; it is not presented
as a universal neural-network stability theorem.

Data-driven unknown-input observers can reproduce model-based observer
solvability conditions from finite data in linear settings
\cite{disaro2025uio}.  Neural simulation relations provide another recent
route for transferring controllers together with probabilistic behavioral
guarantees \cite{nadali2025snsr}.  These works help delimit our claim: we do not
learn an unknown-input observer and do not establish stochastic equivalence
between two black-box systems; we exploit a known sampled identity to remove
the additive disturbance and certify the remaining relative-gain recursion.

Earlier neural identification and control work established that neural
networks can approximate unknown dynamical mappings \cite{narendra1990nn}; our
question is how to choose a control-relevant mapping under asymmetric
information.  Physics-informed machine learning more broadly embeds physical
structure in learning \cite{karniadakis2021piml}, while safe-learning surveys
emphasize explicit constraints and recovery mechanisms \cite{brunke2022safe}.
Simulation-to-real work uses domain randomization \cite{tobin2017domain} and
privileged simulation policies \cite{hwangbo2019skills} to cover variability.
Our deterministic regime grid plays a related coverage role, but the present
evidence remains entirely simulation-based.

\subsection{Position of this paper}
The old ``mechanism versus behavior'' dichotomy is therefore refined into a
composition: privileged behavior supplies diverse, high-performance closed-loop
examples; mechanism chooses a realizable latent target and reconstructs the
action; data fit the remaining history-to-latent map; and a control certificate
states when the composed loop is stable.  This combination, rather than the
choice of multilayer perceptron, is the object studied below.

\begin{table}[t]
\centering
\footnotesize
\setlength{\tabcolsep}{2pt}
\caption{Positioning relative to representative neighboring paradigms.}
\label{tab:positioning}
\begin{tabular}{>{\raggedright\arraybackslash}p{2.1cm}
                >{\raggedright\arraybackslash}p{2.6cm}
                >{\raggedright\arraybackslash}p{2.6cm}
                >{\raggedright\arraybackslash}p{4.1cm}}
\toprule
Paradigm & Training advantage & Learned object & Distinction from this work\\
\midrule
Privileged sensor policy \cite{chen2020learning}
& Simulator semantics/state & Deployable policy & Primarily end-to-end policy transfer\\
History adaptation \cite{kumar2021rma,fu2023deep}
& Environment parameters & Latent plus policy & Empirical adaptation in high-dimensional robotics\\
Parameter-aware expert \cite{paluch2025anc}
& NMPC with true parameters & Recurrent controller & Implicit full-controller imitation; no factorized certificate\\
Realizable-student learning \cite{shenfeld2023tgrl,kim2025realizable}
& Full-state teacher & Student policy/query rule & Focus on information asymmetry, not physical action reconstruction\\
Data-driven UIO \cite{disaro2025uio}
& Finite input--output data & Unknown-input observer & Observer solvability for LTI systems\\
This paper
& Privileged $(a,d)$ expert & Action-derived inverse gain & Exact disturbance elimination, physical reconstruction, and sampled certificate\\
\bottomrule
\end{tabular}
\end{table}

\section{Problem Formulation and Mapping Realizability}

\subsection{Plant and exact sampled timing}
Consider
\begin{equation}
 \dot x_1(t)=a(t)x_2(t)+d(t),\qquad \dot x_2(t)=u(t),
 \label{eq:plant}
\end{equation}
where $x_1$ is regulated, $x_2$ is measured, $a(t)>0$ is an unknown
multiplicative parameter, and $d(t)$ is an unknown additive disturbance.  Let
$r=x_{1\mathrm{ref}}$ and $e=r-x_1$.  With sampling period $T_c$, the verified
Simulink timing is
\begin{align}
 x_{1,k}&=x_{1,k-1}+T_c(a_kx_{2,k-1}+d_k), \label{eq:x1d}\\
 x_{2,k}&=x_{2,k-1}+T_cu_{k-1}. \label{eq:x2d}
\end{align}
Therefore the causal backward difference obeys
\begin{equation}
 \delta x_{1,k}:=\frac{x_{1,k}-x_{1,k-1}}{T_c}
 =a_kx_{2,k-1}+d_k. \label{eq:dxid}
\end{equation}
This timing convention matters: replacing $x_{2,k-1}$ by $x_{2,k}$ destroys
the exact identity used later.

\subsection{Teacher, student, and objective}
The privileged expert sees
$\mathcal I_k^{\rm T}=\{x_{1,k},x_{2,k},r,a_k,d_k\}$ and applies
\begin{align}
 x_{2r,k}&=\frac{k_1e_k-d_k}{a_k}, \label{eq:x2r}\\
 u_{p,k}&=k_2(x_{2r,k}-x_{2,k}), \qquad k_1,k_2>0. \label{eq:up}
\end{align}
If $x_2=x_{2r}$ exactly, then $\dot e=-k_1e$; the privileged variables
select the internal state that cancels $d$ while accounting for $a$.

The deployed student is restricted to
\begin{equation}
 \mathcal I_k^{\rm S}=H_k^L
 =\{x_{1,j},x_{2,j},r_j\}_{j=k-L}^{k}. \label{eq:studentinfo}
\end{equation}
Neither $a$, $d$, nor $x_{2r}$ is a student input.  Privileged variables may
be used to construct the expert and audit a dataset, but the recommended
student label in Section 5 also avoids the true $a$.

Given expert rollouts over a declared regime set $\Omega$, find a causal
controller $u_k=\pi_\theta(H_k^L)$ that approaches expert tracking on unseen
trajectories in $\Omega$ and admits an explicit stability statement.  This is
not a claim that classical robust/adaptive control is impossible; it is a test
of whether offline privileged multi-regime knowledge can improve one fixed
deployable design.

\subsection{Why a nominal disturbance observer does not remove the gain problem}
The simulation baseline uses the nominal-model disturbance observer
\begin{align}
 x_{2r}&=k_1e-\hat d, & u&=k_2(x_{2r}-x_2), \label{eq:dobcontrol}\\
 \hat d&=L(x_1-p), & \dot p&=x_2+\hat d. \label{eq:dob}
\end{align}
This design is exact for the nominal input coefficient $a=1$.  For $a\ne1$,
rewrite the actual plant as
\begin{equation}
 \dot x_1=x_2+\bar d,\qquad
 \bar d=(a-1)x_2+d. \label{eq:lumpedd}
\end{equation}
Thus the observer is not estimating $d$ alone; it must track a lumped quantity
that contains the controlled state.  For constant $a$, differentiating
\eqref{eq:dob} and defining $\tilde d=\bar d-\hat d$ gives
\begin{align}
 \dot{\hat d}&=L(\bar d-\hat d)=L\tilde d,\\
 \dot{\tilde d}&=(a-1)u+\dot d-L\tilde d. \label{eq:doberror}
\end{align}
When $a=1$ and $d$ is constant, the forcing term vanishes and increasing $L$
accelerates estimation.  When $a\ne1$, the control action itself drives the
estimation error through $(a-1)u$.  The observer, inner controller, and unknown
gain are therefore dynamically coupled.  A larger $L$ can reduce error in a
moderate regime but also increases noise/control demand and, after sampling,
can approach a stability boundary.  Equation \eqref{eq:doberror} does not imply
that all advanced observers or adaptive controllers must fail; it explains why
one fixed nominal DOB is not automatically a solution to simultaneous gain and
disturbance uncertainty.  Section 7 evaluates this mechanism separately over
$a<1$, $a=1$, and $a>1$.

\subsection{Three levels of realizability}
\begin{definition}[Instantaneous action realizability]
The expert is instantaneously realizable on a dataset if a single-valued map
$\pi$ satisfies $u_{p,k}=\pi(x_{1,k},x_{2,k},r_k)$ within a declared tolerance.
\end{definition}

\begin{proposition}[State--action conflict]
If two samples have identical student observations and different expert
actions, no deterministic instantaneous student reproduces both.
\end{proposition}
\begin{proof}
For identical observations $o_i=o_j$, a function must satisfy
$\pi(o_i)=\pi(o_j)$.  This contradicts $u_{p,i}\ne u_{p,j}$.
\end{proof}

With continuous data, exact collisions are rare.  We therefore normalize each
observable coordinate, search cross-regime nearest neighbors, and examine
\begin{equation}
 \mathcal C_\varepsilon=
 \{(i,j):\|o_i-o_j\|\le\varepsilon,
       |u_{p,i}-u_{p,j}|>\tau_u\}. \label{eq:conflictset}
\end{equation}
A large conflict set or large local conditional variance
$\operatorname{Var}(u_p\mid o)$ indicates an ill-conditioned static map.

\begin{definition}[History realizability]
The expert is history-realizable over length $L$ if a causal map of
$H_k^L$ reproduces the relevant expert quantity.  History helps only when
different hidden regimes leave distinguishable observable trajectories.
\end{definition}

\begin{definition}[Task realizability]
A student is task-realizable if it need not reproduce every privileged
internal variable or action exactly, but the component of its error that enters
the regulated closed loop can be bounded sufficiently for the desired task.
\end{definition}

These notions suggest a sequence.  Diagnose static conflicts; introduce
history only if hidden variables affect observable dynamics; then use plant and
expert structure to identify the smallest task-relevant latent variable.  The
next section performs the last step exactly for \eqref{eq:plant}.

\section{Exact Mechanism-Guided Task Reduction}

\subsection{Eliminating the additive disturbance}
Set $\alpha_k=1/a_k$.  Equation \eqref{eq:dxid} gives
\begin{equation}
 d_k=\delta x_{1,k}-a_kx_{2,k-1}. \label{eq:dreplace}
\end{equation}
Substitute \eqref{eq:dreplace} into the privileged internal reference:
\begin{align}
 x_{2r,k}
 &=\alpha_k\{k_1e_k-[\delta x_{1,k}-a_kx_{2,k-1}]\}\nonumber\\
 &=\alpha_k(k_1e_k-\delta x_{1,k})+x_{2,k-1}. \label{eq:x2rfact}
\end{align}
Subtract $x_{2,k}$ and multiply by $k_2$.

\begin{proposition}[Exact expert-action factorization]
Under the sampled timing \eqref{eq:x1d}--\eqref{eq:x2d},
\begin{equation}
 u_{p,k}=k_2\left[
 \alpha_k(k_1e_k-\delta x_{1,k})+x_{2,k-1}-x_{2,k}
 \right]. \label{eq:factor}
\end{equation}
\end{proposition}
\begin{proof}
Equation \eqref{eq:x2rfact} inserted into \eqref{eq:up} gives
\eqref{eq:factor} directly.
\end{proof}

Define the measurable task coordinate and sampled correction
\begin{equation}
 q_k=k_1e_k-\delta x_{1,k},\qquad
 \Delta x_{2,k}=x_{2,k-1}-x_{2,k}. \label{eq:qdelta}
\end{equation}
Then $u_{p,k}=k_2(\alpha_kq_k+\Delta x_{2,k})$.  The large unknown $d_k$
has disappeared; it is already encoded in the measured derivative.  The only
unmeasured factor needed to reconstruct the expert action is $\alpha_k$.

\subsection{What is actually learned?}
The deployed structured controller is
\begin{equation}
 u_k=\sat_{u_{\max}}\left\{
 k_2[\hat\alpha_kq_k+\Delta x_{2,k}]
 \right\}. \label{eq:studentlaw}
\end{equation}
It is correct to interpret the neural module as an inverse-gain identifier.
However, this does not reduce the method to imitation of a conventional
observer:
\begin{enumerate}
\item $d$ is not separately identified; it is removed by the sampled physical
identity;
\item the latent target is selected by action sensitivity, not by a demand to
reconstruct the entire unknown plant;
\item training trajectories are generated by a privileged closed-loop expert
over deliberately varied regimes, rather than by copying an observer;
\item the final control action is a transparent composition of learned
adaptation and known feedback.
\end{enumerate}

The action error caused only by latent estimation is
\begin{equation}
 \tilde u_{\alpha,k}=u_k-u_{p,k}
 =k_2q_k(\hat\alpha_k-\alpha_k) \label{eq:actionerr}
\end{equation}
before saturation and measurement errors.  Hence an inaccurate parameter
estimate near $q_k=0$ can be harmless, whereas the same error during a large
transient can matter strongly.  This is why parameter RMSE and uniform
sample weighting are not aligned automatically with the control task.

\subsection{Why history can reveal the gain}
At one instant, \eqref{eq:dxid} gives one equation in two unknowns $a_k,d_k$;
there is no unique algebraic separation.  Over a window in which $a,d$ are
approximately constant,
\begin{equation}
 \delta x_{1,j}=a x_{2,j-1}+d,\qquad j=k-L+1,\ldots,k. \label{eq:windowreg}
\end{equation}
If $x_2$ varies, the slope of this affine relation contains $a$ and the
intercept contains $d$.  In the noiseless two-sample case,
\begin{equation}
 a=\frac{\delta x_{1,i}-\delta x_{1,j}}
          {x_{2,i-1}-x_{2,j-1}}, \qquad x_{2,i-1}\ne x_{2,j-1}. \label{eq:twosample}
\end{equation}
Equation \eqref{eq:twosample} is not the implemented estimator, but it explains
the information mechanism: excitation and temporal variation make the gain
identifiable.  Near equilibrium, weak variation makes raw identification
ill-conditioned; fortunately $|q_k|$ and therefore the action sensitivity in
\eqref{eq:actionerr} are also often small.  The learned estimator captures a
regularized, multi-sample version of this relation.

\begin{figure}[t]
\centering
\includegraphics[width=\textwidth]{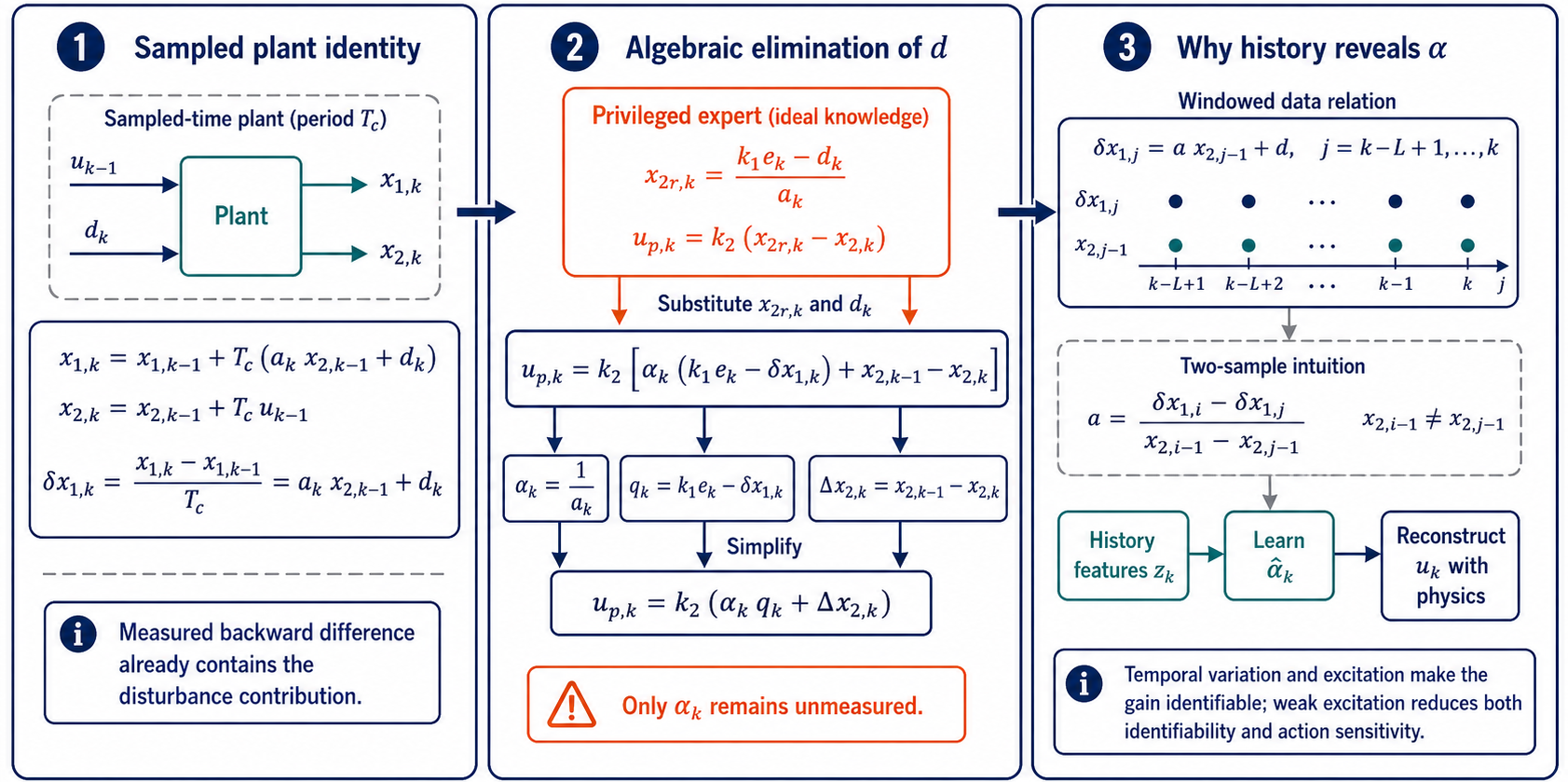}
\caption{The process of the proposed method.}
\label{fig:task_reduction_infographic}
\end{figure}

Figure~\ref{fig:task_reduction_infographic} collects the three steps of this
argument---sampled-data identity, algebraic elimination of $d$, and gain
inference from causal history---and shows where the learned component enters
the otherwise fixed physical reconstruction.

\section{Learning Design and Privileged Sample Use}

\subsection{Separation of three data roles}
To avoid ambiguity, we distinguish three interfaces.
\begin{enumerate}
\item \emph{Expert-generation interface:} $(x_1,x_2,r,a,d)$ are available to
\eqref{eq:x2r}--\eqref{eq:up}.  This is where privileged information is used.
\item \emph{Student-training record:} expert actions $u_p$ and measured
trajectories are retained.  The true $a,d$ may be kept only as audit metadata.
\item \emph{Deployment interface:} only the current/past $(x_1,x_2,r)$ enter
the estimator and \eqref{eq:studentlaw}.
\end{enumerate}
Thus the dataset is \emph{generated by privileged control} even though the
recommended regression columns and deployed controller contain no privileged
variable.  Calling it ``non-privileged data collection'' would be misleading.

\subsection{Causal features}
Let the sparse delays be
\begin{equation}
 \mathcal L=\{0,1,5,10,50,100,200,500,1000\}.
\end{equation}
The estimator input is
\begin{equation}
 z_k=\col_{j\in\mathcal L}
 (e_{k-j},x_{2,k-j},\delta x_{1,k-j})\in\R^{27}. \label{eq:features}
\end{equation}
At $T_c=10^{-4}$ s, this covers 100 ms while retaining dense recent taps.
Training-set percentiles scale each coordinate to approximately $[-1,1]$;
values are clipped at deployment.  Histories never cross trajectory
boundaries.  Complete trajectories, not individual samples, are assigned to
train, validation, and test sets.

\subsection{Action-derived supervision}
The direct privileged label would be $\alpha_k=1/a_k$.  A stronger separation
between expert privilege and student supervision follows by rearranging
\eqref{eq:factor}.  Whenever $|q_k|\ge q_{\min}$,
\begin{equation}
 \alpha_k^u=
 \frac{u_{p,k}/k_2-\Delta x_{2,k}}{q_k}. \label{eq:alphau}
\end{equation}
This label uses the expert's action and deployment-visible trajectory only.  It
does not claim that $a$ was unknown when the expert samples were generated;
rather, it shows that the student need not be trained to reproduce a
privileged parameter explicitly.

Division by small $q_k$ is ill-conditioned.  Such samples are removed from the
latent regression and the remaining samples receive task weight
\begin{equation}
 \omega_k=c_0+\min\left(\frac{|q_k|}{q_{\rm cap}},1\right)^2, \qquad
 \frac{1}{N}\sum_k\omega_k=1. \label{eq:weight}
\end{equation}
The nonzero $c_0$ preserves background coverage; the second term emphasizes
samples for which latent error creates a large action error.

\subsection{Estimator and training objective}
A feedforward network $f_\theta:\R^{27}\rightarrow\R$ has hidden widths 64
and 32 with ReLU activations.  It predicts a log inverse gain,
\begin{equation}
 \hat\alpha_k=\exp\{\operatorname{clip}[
 f_\theta(z_k),\log\alpha_L,\log\alpha_U]\},
 \quad (\alpha_L,\alpha_U)=(0.5,12). \label{eq:alphaout}
\end{equation}
The log parameterization enforces positivity; clipping bounds extrapolation.
The loss is
\begin{equation}
 \mathcal J(\theta)=\frac{1}{N}\sum_{k=1}^{N}
 \omega_k[\log\hat\alpha_k-\log\alpha_k^u]^2
 +\lambda\|\theta\|_2^2. \label{eq:loss}
\end{equation}
Validation selects the checkpoint.  Closed-loop validation remains mandatory
because \eqref{eq:loss} is a one-step supervised objective.

\subsection{Algorithmic route}
The complete route is summarized without committing to a particular drawing:
\begin{enumerate}
\item declare the operating domain $\Omega$ and sampling/timing convention;
\item generate multi-regime closed-loop trajectories with the privileged
expert;
\item test instantaneous state--action conflicts and history informativeness;
\item derive a physically composable latent target and action layer;
\item construct action-derived labels, reject ill-conditioned divisions, and
split by trajectories;
\item train the latent estimator and verify both offline error and closed-loop
rollouts;
\item certify a relative-gain interval and report all trajectories leaving the
certificate or reaching saturation.
\end{enumerate}

\begin{remark}
The method is end-to-end in its deployment interface---measured histories map
to $u$---but structured internally.  ``End-to-end'' here does not mean that the
entire control law is an unconstrained neural network.
\end{remark}

\section{Detailed Sampled-Data Stability Analysis}

\subsection{Step 1: choose coordinates that expose cancellation}
Analyze an unsaturated interval on which $a,d$ are constant.  Define
\begin{equation}
 w_k=ax_{2,k}+d,\qquad \rho_k=a\hat\alpha_k. \label{eq:wrho}
\end{equation}
The coordinate $w$ is the part of $\dot x_1$ produced by the internal state
and disturbance.  From \eqref{eq:dxid}, $\delta x_{1,k}=w_{k-1}$.  Because
$e_k=r-x_{1,k}$ and $r$ is constant,
\begin{equation}
 e_{k+1}=e_k-T_cw_k. \label{eq:erec}
\end{equation}

\subsection{Step 2: rewrite the implemented action}
Ignoring saturation, multiply \eqref{eq:studentlaw} by $a$:
\begin{align}
 au_k
 &=k_2[a\hat\alpha_k(k_1e_k-w_{k-1})
       +a(x_{2,k-1}-x_{2,k})] \nonumber\\
 &=k_2[\rho_kk_1e_k-\rho_kw_{k-1}+w_{k-1}-w_k] \nonumber\\
 &=k_2[\rho_kk_1e_k+(1-\rho_k)w_{k-1}-w_k]. \label{eq:au}
\end{align}
The constant $d$ cancels in the difference
$a(x_{2,k-1}-x_{2,k})=w_{k-1}-w_k$.  Moreover,
\begin{equation}
 w_{k+1}=w_k+T_cau_k. \label{eq:wrec}
\end{equation}

\subsection{Step 3: obtain the exact augmented recursion}
The backward difference introduces one-sample memory, so the correct state is
$\xi_k=[e_k,w_k,w_{k-1}]^T$.  Equations
\eqref{eq:erec}, \eqref{eq:au}, and \eqref{eq:wrec} give
\begin{equation}
 \xi_{k+1}=A_d(\rho_k)\xi_k, \label{eq:augrec}
\end{equation}
with
\begin{equation}
A_d(\rho)=
\begin{bmatrix}
1&-T_c&0\\
T_ck_1k_2\rho&1-T_ck_2&T_ck_2(1-\rho)\\
0&1&0
\end{bmatrix}. \label{eq:Ad}
\end{equation}
All unknown-gain and learned-estimator effects enter the homogeneous dynamics
through the dimensionless relative inverse gain $\rho=a\hat\alpha$.  Exact
estimation corresponds to $\rho=1$, but stability need not require equality.

\subsection{Step 4: reduce infinitely many checks to two endpoints}
\begin{lemma}[Matrix convexity]
For fixed $P\succ0$, $A_d(\rho)^TPA_d(\rho)$ is matrix convex in $\rho$.
\end{lemma}
\begin{proof}
Let $A_i=A_d(\rho_i)$ and
$A_\lambda=\lambda A_1+(1-\lambda)A_2$.  Direct expansion gives
\begin{align}
&\lambda A_1^TPA_1+(1-\lambda)A_2^TPA_2
-A_\lambda^TPA_\lambda\nonumber\\
&\quad=\lambda(1-\lambda)(A_1-A_2)^TP(A_1-A_2)\succeq0.
\end{align}
Since $A_d$ is affine in $\rho$, the result follows.
\end{proof}

\begin{theorem}[Uniform exponential stability]
Suppose the actuator is unsaturated, $a,d$ are constant on the analyzed
interval, and $\rho_k\in[\rho_L,\rho_U]$.  If $P\succ0,Q\succ0$ satisfy
\begin{equation}
 A_d(\rho_i)^TPA_d(\rho_i)-P\preceq-Q,
 \quad \rho_i\in\{\rho_L,\rho_U\}, \label{eq:lmi}
\end{equation}
then the origin of \eqref{eq:augrec} is uniformly exponentially stable for
every time-varying sequence $\rho_k$ in the interval.
\end{theorem}
\begin{proof}
Any $\rho\in[\rho_L,\rho_U]$ is a convex combination of the endpoints.
Matrix convexity extends \eqref{eq:lmi} to the whole interval.  With
$V_k=\xi_k^TP\xi_k$,
\begin{equation}
 V_{k+1}-V_k\le-\xi_k^TQ\xi_k
 \le-\frac{\lambda_{\min}(Q)}{\lambda_{\max}(P)}V_k.
\end{equation}
Therefore $V_k\le\eta^{k-k_0}V_{k_0}$ for
$\eta=1-\lambda_{\min}(Q)/\lambda_{\max}(P)\in(0,1)$, and
\begin{equation}
 \|\xi_k\|_2\le
 \sqrt{\frac{\lambda_{\max}(P)}{\lambda_{\min}(P)}}
 \eta^{(k-k_0)/2}\|\xi_{k_0}\|_2.
\end{equation}
\end{proof}

\subsection{Step 5: bounded residuals imply practical stability}
Let differentiation, modeling, and action residuals enter as
\begin{equation}
 \xi_{k+1}=A_d(\rho_k)\xi_k+Bg_k,\qquad B=[0,1,0]^T. \label{eq:forced}
\end{equation}
For a derivative error $\nu_k$ and additive action error
$\varepsilon_{u,k}$,
\begin{equation}
 g_k=T_c[-k_2\rho_k\nu_k+a\varepsilon_{u,k}]. \label{eq:g}
\end{equation}
Assume the stronger contraction
$A_d(\rho)^TPA_d(\rho)\preceq\gamma^2P$, $0<\gamma<1$.  Using
$\|x\|_P=\sqrt{x^TPx}$ and the induced norm,
\begin{align}
 \|\xi_{k+1}\|_P
 &\le\gamma\|\xi_k\|_P+\|B\|_P|g_k|,\\
 \|\xi_k\|_P
 &\le\gamma^{k-k_0}\|\xi_{k_0}\|_P
 +\|B\|_P\sum_{j=k_0}^{k-1}\gamma^{k-1-j}|g_j|.
\end{align}
If $|g_k|\le\bar g$, then
\begin{equation}
 \limsup_{k\to\infty}\|\xi_k\|_2
 \le\frac{\sqrt{B^TPB}}
 {\sqrt{\lambda_{\min}(P)}(1-\gamma)}\bar g. \label{eq:iss}
\end{equation}
This is a sufficient worst-case bound, not a tight RMSE prediction.

\subsection{Regime changes and saturation}
If $a,d$ jump, the coordinate $w=ax_2+d$ jumps even when the physical state is
continuous:
\begin{equation}
 \Delta w_k=x_{2,k}\Delta a_k+\Delta d_k. \label{eq:jump}
\end{equation}
Thus within-regime stability and post-switch recovery are distinct.  A
dwell-time argument can combine \eqref{eq:iss} and
\eqref{eq:jump}; arbitrary infinitely fast changes are not covered.  Actuator
saturation also breaks the homogeneous recursion and is reported empirically.

\subsection{Numerical certificate}
For $k_1=10$, $k_2=100$, and $T_c=10^{-4}$ s, endpoint LMIs are feasible for
\begin{equation}
 \rho=a\hat\alpha\in[0.1,4]. \label{eq:certinterval}
\end{equation}
A reproducible solution is
\begin{equation}
P=10^6\!\begin{bmatrix}
197.4643&-10.0837&0.3732\\
-10.0837&14.7660&-13.5732\\
0.3732&-13.5732&13.5716
\end{bmatrix}. \label{eq:Pnum}
\end{equation}
Its minimum eigenvalue is $3.5215\times10^5$.  The minimum eigenvalues of
$P-A_d^TPA_d$ at $\rho=0.1$ and $4$ are 200.82 and 670.74; the computed
contraction factor is $\gamma=0.99983295$.  On the independent original-timing
dataset, 99.9312\% of valid histories fall in this interval.  This empirical
coverage supports relevance but does not replace the explicit assumption
$\rho_k\in[0.1,4]$.

\section{Simulation Study}

\subsection{Protocol and hypotheses}
Simulations use $k_1=10$, $k_2=100$, $T_c=10^{-4}$ s, $r=1$, and the
27-feature 64--32 ReLU estimator described above.  The action is limited to
$\pm12000$.  Complete trajectories define all splits.  Tracking metrics omit
the first second unless a transition window is explicitly stated.  Every
controller receives the same parameter and disturbance sequence in a given
comparison.

We test five hypotheses: (H1) one nominal disturbance observer exhibits an
asymmetric parameter-dependent limitation under simultaneous uncertainty;
(H2) direct action imitation is not reliably closed-loop realizable; (H3)
action-derived latent supervision is sufficient; (H4) the structured student
approaches the privileged expert over simultaneous $a,d$ changes; and (H5) its
advantage is not an artifact of an obviously weak observer tuning.

The baselines are the privileged expert, a DOBC/observer controller, a static
direct-action network, and a history direct-action network.  The observer gain
is swept independently on four unseen tuning seeds.  The best tested
non-failing gain, $L=19000$, is frozen for final comparisons.  This is a fair
empirical protocol, not a global optimality claim for all observers.

\subsection{Problem evidence: observer behavior across parameter regimes}
\begin{figure}[t]
\centering
\includegraphics[width=\textwidth]{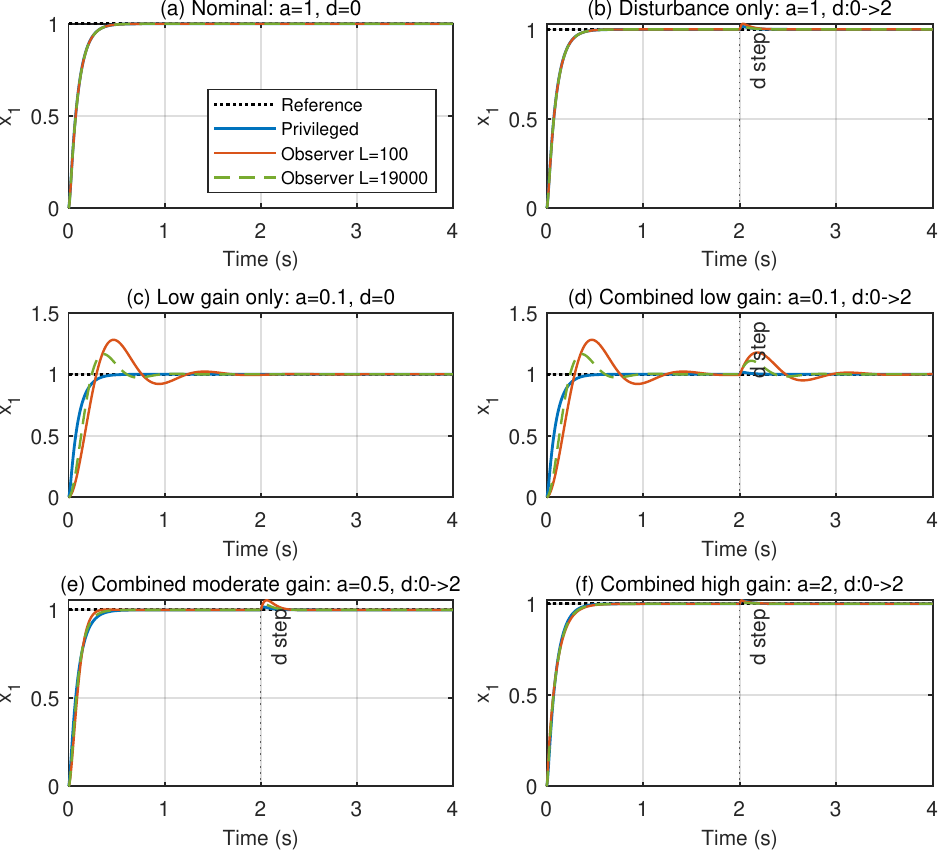}
\caption{Canonical observer scenarios.  The nominal and $a>1$ cases are
benign, whereas $a=0.1$ causes large parameter-induced startup transients that
remain present when the disturbance step is added.  A higher observer gain
reduces but does not remove this effect.  Disturbance cases use
$d:0\rightarrow2$ at $t=2$ s over a 4-s run.}
\label{fig:observer_canonical}
\end{figure}

Before evaluating the student, we isolate the difficulty that motivates
additional adaptation.  The legacy simulation cases are reproduced and
extended: nominal $(a,d)=(1,0)$, disturbance only $(1,0\rightarrow2)$,
parameter mismatch only $(0.1,0)$, and combined uncertainty
$(0.1,0\rightarrow2)$.  Two further combined cases, $a=0.5$ and $a=2$, expose
the difference between weakened and strengthened plant channels.  The
privileged expert is compared with the original observer gain $L=100$ and the
independently tuned gain $L=19000$ in Fig.~\ref{fig:observer_canonical}.  For
the four disturbance cases, the step occurs at $t=2$ s and each run ends at
$t=4$ s; no disturbance-step marker is drawn for the nominal and parameter-only
cases.

\begin{figure}[t]
\centering
\includegraphics[width=\textwidth]{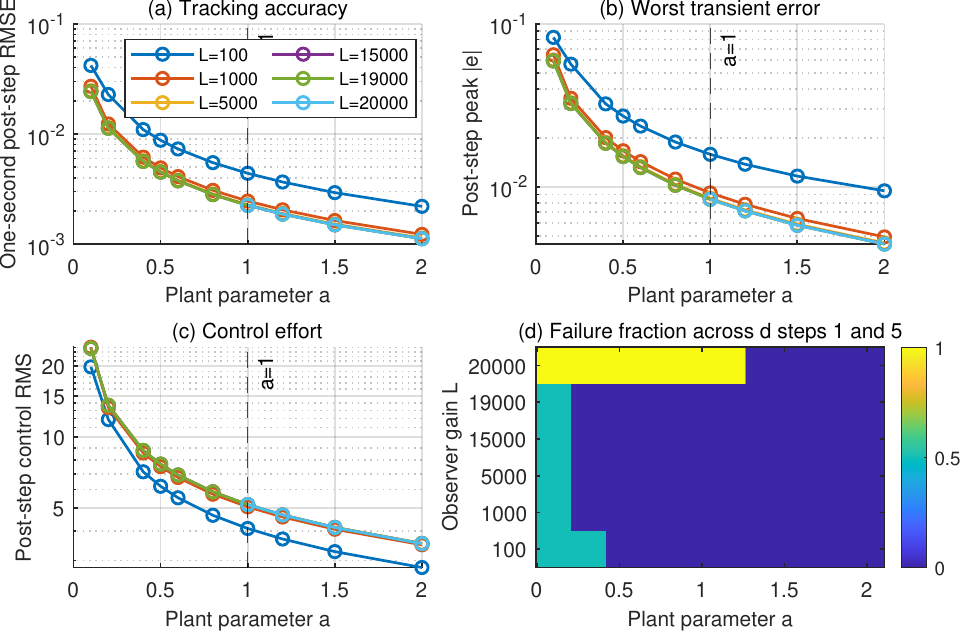}
\caption{Systematic observer scan across $a<1$, $a=1$, and $a>1$ for a unit
disturbance step.  Failed cases are omitted from the continuous metric curves
and retained in the failure map.  The high-gain boundary is confined to the
weakened-channel region in this experiment.}
\label{fig:observer_regimes}
\end{figure}

To separate the canonical visualization from the broader robustness diagnosis,
the systematic scan below retains the unit-scale disturbance grid used in the
original diagnostic dataset.  We then sweep
\begin{equation}
a\in\{0.1,0.2,0.4,0.5,0.6,0.8,1,1.2,1.5,2\},
\quad d:0\rightarrow\{0,1,5\},
\end{equation}
and $L\in\{100,1000,5000,15000,19000,20000\}$.  The one-second
post-step metrics in Table~\ref{tab:observer_regimes} and
Fig.~\ref{fig:observer_regimes} support three observations.

First, at fixed disturbance magnitude, observer error decreases as $a$
increases.  For $d:0\rightarrow1$ and $L=19000$, RMSE falls from 0.02423 at
$a=0.1$ to 0.002246 at $a=1$ and 0.001119 at $a=2$.  The difficulty is thus
strongly asymmetric: a small $a$ weakens the effective plant channel while the
nominal observer/controller still acts as if $a=1$.  Second, increasing $L$
has a cost.  At $a=0.1,d:0\rightarrow1$, moving from $L=100$ to $19000$
reduces RMSE from 0.04185 to 0.02423, but increases control RMS from 19.87 to
23.81 and peak control from 57.92 to 110.17.  Third, the next tested gain
$L=20000$ diverges for every $a<1$ but remains stable for all tested $a\ge1$.
This parameter-dependent stability boundary explains why tuning on the nominal
case alone is misleading.

\begin{table}[t]
\centering
\small
\caption{Observer one-second post-disturbance RMSE over representative regimes.
An asterisk denotes the common RMSE-based task-failure criterion; ``div.''
denotes numerical divergence.}
\label{tab:observer_regimes}
\begin{tabular}{ccccc}
\toprule
$a$ & $d$ step & $L=100$ & $L=19000$ & $L=20000$\\
\midrule
0.1 & $0\to1$ & 0.04185 & 0.02423 & div.\\
0.1 & $0\to5$ & 0.21130$^*$ & 0.11416$^*$ & div.\\
0.5 & $0\to1$ & 0.00876 & 0.00450 & div.\\
1.0 & $0\to1$ & 0.00439 & 0.00225 & 0.00225\\
2.0 & $0\to1$ & 0.00220 & 0.00112 & 0.00112\\
\bottomrule
\end{tabular}
\end{table}

These results do not prove the necessity of neural control in a universal
sense; gain scheduling, adaptive control, or a redesigned observer are possible
alternatives.  They establish the narrower motivation needed here: a single
nominal DOB does not reproduce regime-aware performance across the declared
gain--disturbance domain, and aggressive tuning cannot fix the gap uniformly.
The following experiments test whether privileged multi-regime data can supply
the missing adaptation while preserving a fixed deployable interface.

\subsection{Complete closed-loop signals in the canonical regimes}
\begin{figure}[t]
\centering
\includegraphics[width=\textwidth]{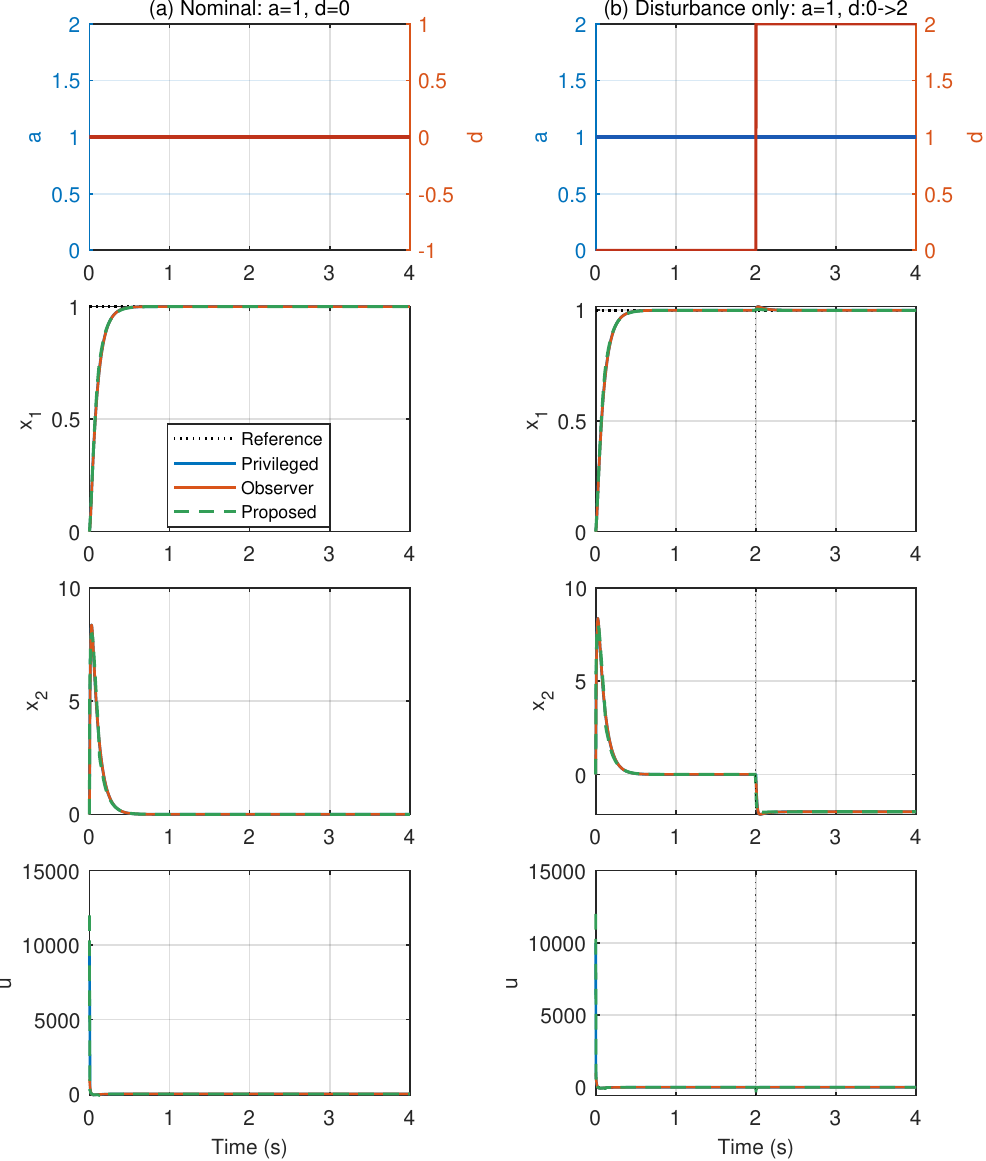}
\caption{Complete closed-loop signals for the nominal and disturbance-only
regimes.  The disturbance changes from 0 to 2 at $t=2$ s in the right column.}
\label{fig:full_nominal_disturbance}
\end{figure}

\begin{figure}[t]
\centering
\includegraphics[width=\textwidth]{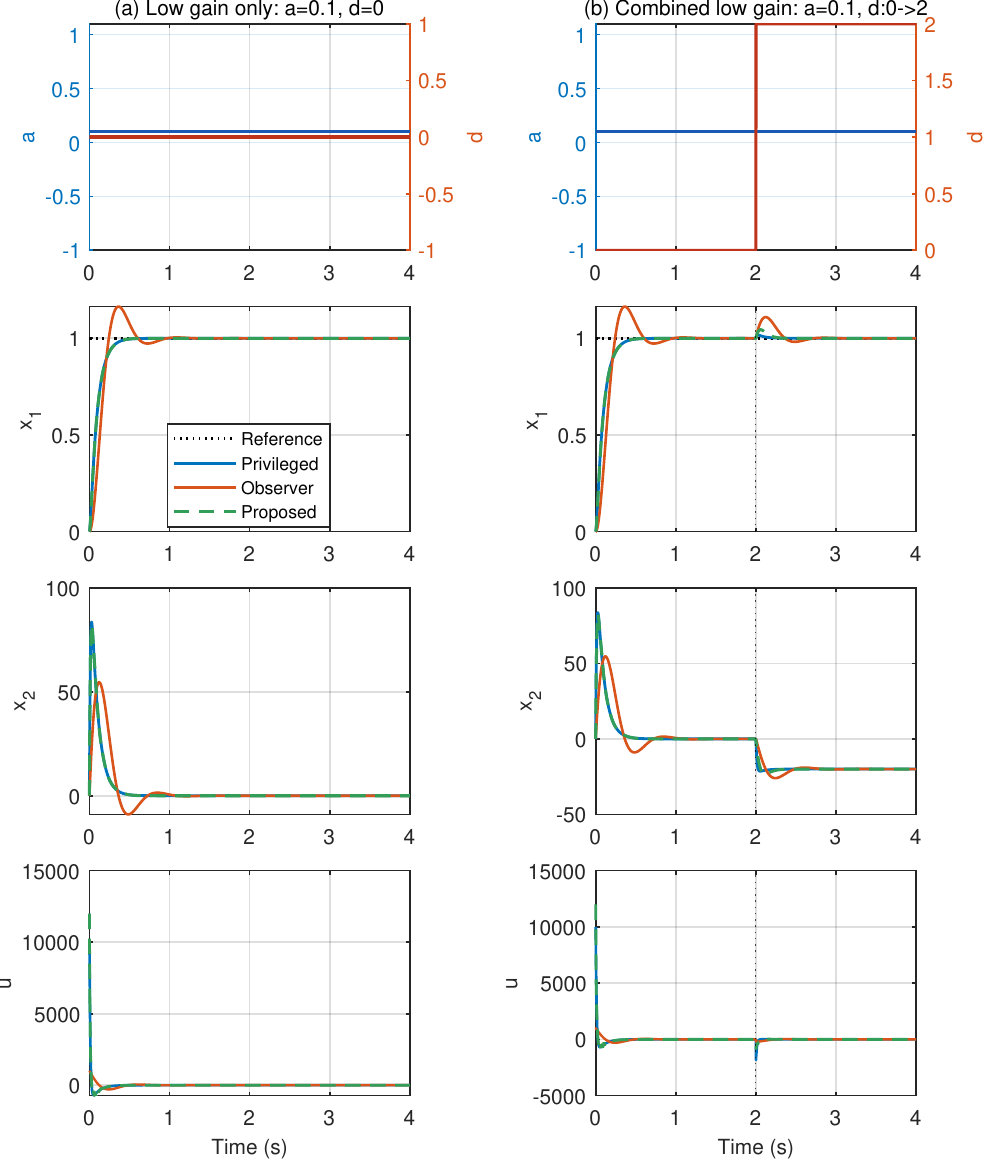}
\caption{Complete closed-loop signals for the low-gain parameter-only and
combined low-gain regimes.  All three controllers act on independent plant
copies subject to the same $a$ and $d$.}
\label{fig:full_low_gain}
\end{figure}

\begin{figure}[t]
\centering
\includegraphics[width=\textwidth]{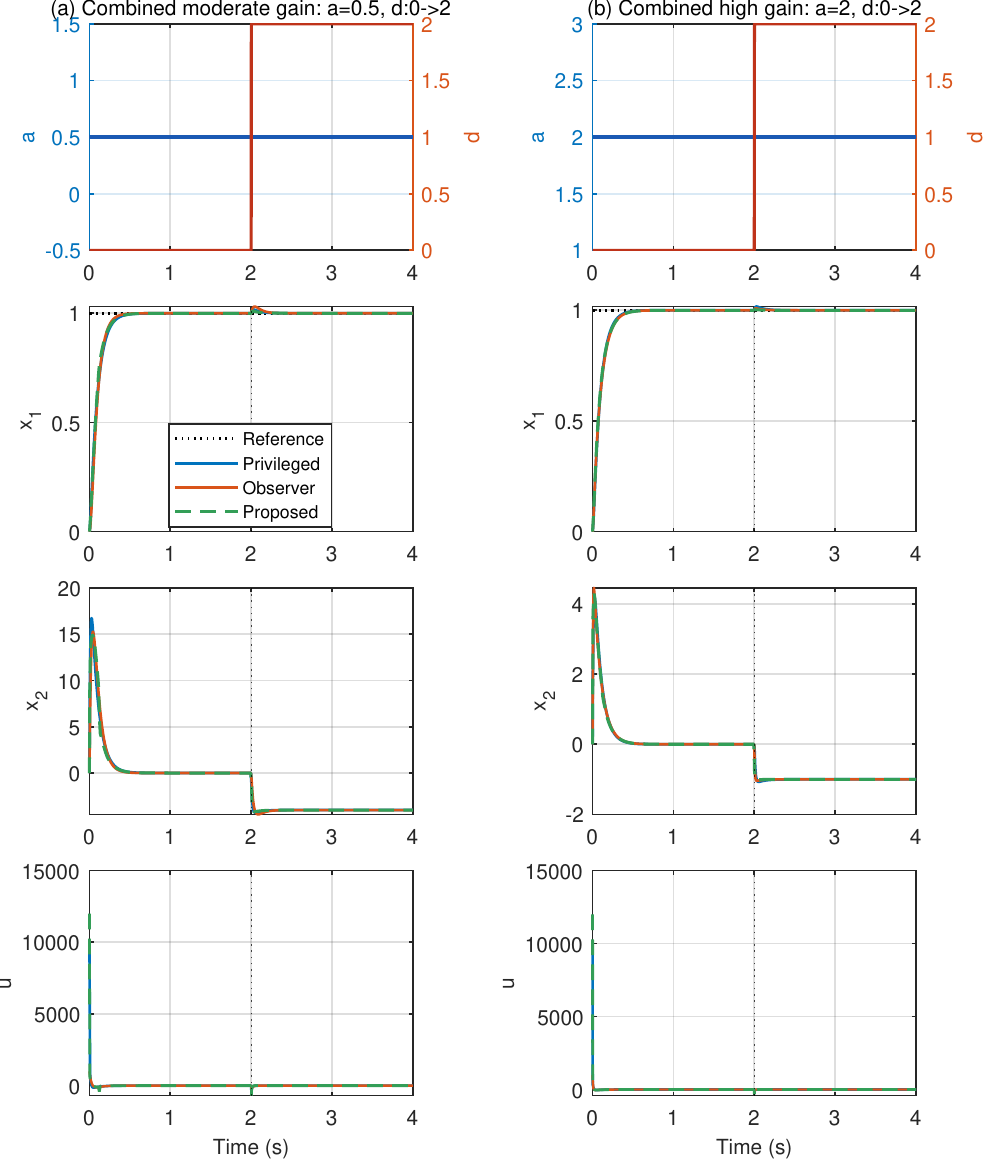}
\caption{Complete closed-loop signals for the combined moderate- and
high-gain regimes.}
\label{fig:full_moderate_high}
\end{figure}

The preceding observer-only diagnosis is complemented by a three-controller
comparison in the same Simulink model.  Each controller drives its own copy of
the plant, so the plotted $x_2$ trajectories are recorded separately rather
than inferred from another loop.  Figures~\ref{fig:full_nominal_disturbance}--
\ref{fig:full_moderate_high} report the actual operating-condition signals
$(a,d)$ together with $x_1$, $x_2$, and $u$ for the privileged expert, tuned
observer, and proposed student.  The tuned observer uses the frozen
$L=19000$ selected independently above.

For a consistent comparison, Table~\ref{tab:canonical_complete_metrics}
evaluates every case over $t\in[2,4)$ s.  This interval is the post-step window
for disturbance cases and a settled-operation window otherwise.  The proposed
student reduces $x_1$ RMSE relative to the tuned observer by approximately
69\%, 72\%, and 72\% in the combined $a=0.1$, $0.5$, and $2$ cases,
respectively.  This improvement is not free: its control RMS and peak are
usually larger than those of the observer and closer to the scale of the
privileged expert.  The $x_2$ RMS values near 20, 4, and 1 in these three
regimes are consistent with the equilibrium compensation magnitude $d/a$.

\begin{table}[t]
\centering
\scriptsize
\setlength{\tabcolsep}{2.7pt}
\caption{Canonical-regime metrics on $t\in[2,4)$ s.  N: nominal; D:
disturbance only; G: low-gain parameter only; LG/MG/HG: combined low,
moderate, and high gain.  P, O, and S denote privileged, tuned observer, and
proposed student.  Full-precision values are released in the accompanying CSV.}
\label{tab:canonical_complete_metrics}
\begin{tabular}{llrrrrr}
\toprule
Case & Ctrl. & RMSE$(e_1)$ & $\max|e_1|$ & RMS$(x_2)$ & RMS$(u)$ & $\max|u|$\\
\midrule
N  & P & $2.71\mathrm{e}{-11}$ & $1.82\mathrm{e}{-10}$ & $3.05\mathrm{e}{-10}$ & $3.44\mathrm{e}{-9}$ & $2.31\mathrm{e}{-8}$\\
N  & O & $2.74\mathrm{e}{-11}$ & $1.84\mathrm{e}{-10}$ & $3.09\mathrm{e}{-10}$ & $3.48\mathrm{e}{-9}$ & $2.33\mathrm{e}{-8}$\\
N  & S & $1.62\mathrm{e}{-10}$ & $1.04\mathrm{e}{-9}$  & $1.67\mathrm{e}{-9}$  & $1.71\mathrm{e}{-8}$ & $1.10\mathrm{e}{-7}$\\
D  & P & 0.00320 & 0.01688 & 2.002 & 10.21 & 183.96\\
D  & O & 0.00318 & 0.01680 & 2.002 & 10.21 & 187.27\\
D  & S & 0.00088 & 0.00516 & 2.000 & 18.01 & 579.34\\
G  & P & $2.74\mathrm{e}{-11}$ & $1.84\mathrm{e}{-10}$ & $3.09\mathrm{e}{-9}$ & $3.48\mathrm{e}{-8}$ & $2.33\mathrm{e}{-7}$\\
G  & O & $6.20\mathrm{e}{-6}$ & $2.64\mathrm{e}{-5}$ & $9.68\mathrm{e}{-4}$ & 0.00981 & 0.03564\\
G  & S & $5.02\mathrm{e}{-15}$ & $1.38\mathrm{e}{-14}$ & $6.44\mathrm{e}{-12}$ & $5.22\mathrm{e}{-11}$ & $8.47\mathrm{e}{-10}$\\
LG & P & 0.00320 & 0.01688 & 20.021 & 102.12 & 1839.6\\
LG & O & 0.03180 & 0.10975 & 20.246 & 44.61 & 203.17\\
LG & S & 0.00992 & 0.04462 & 20.072 & 65.46 & 894.12\\
MG & P & 0.00320 & 0.01688 & 4.004 & 20.42 & 367.92\\
MG & O & 0.00636 & 0.03067 & 4.009 & 15.29 & 195.93\\
MG & S & 0.00178 & 0.00982 & 4.002 & 27.10 & 696.68\\
HG & P & 0.00320 & 0.01688 & 1.001 & 5.11 & 91.98\\
HG & O & 0.00159 & 0.00897 & 1.000 & 6.86 & 171.30\\
HG & S & 0.00045 & 0.00272 & 1.000 & 11.53 & 420.74\\
\bottomrule
\end{tabular}
\end{table}

\subsection{Offline fit does not predict closed-loop success}
\begin{table}[t]
\centering
\caption{Structural and supervision ablation.  ``Failure'' uses the common
task-failure rule in the released scripts.}
\label{tab:ablation}
\begin{tabular}{lcc}
\toprule
Method & Action NRMSE & Closed-loop result\\
\midrule
Static direct action & 0.3960 & 4/4 failures\\
History direct action & 0.3451 & 4/4 failures\\
Structured, uniform $\alpha$ & 0.2053 & RMSE 0.01101\\
Structured, action-derived & 0.1746 & RMSE 0.01159\\
Structured, task-weighted $\alpha$ & 0.1611 & RMSE 0.01279\\
\bottomrule
\end{tabular}
\end{table}

Table \ref{tab:ablation} separates supervised action fit from control.  Adding
history improves the direct network's NRMSE but both direct policies fail every
closed-loop trial.  Structured students remain stable and close to the expert.
The action-derived version never uses true $a$ as a label and is selected as
the main method.  The lowest offline action error does not yield the lowest
closed-loop RMSE, reinforcing that the learned target and rollout distribution
matter more than a single regression score.

\subsection{Deterministic parameter--disturbance grid}
The student comparison is intentionally restricted to the training-covered and
near-boundary gain domain (the original-timing dataset has
$a\in[0.1,0.864]$, while the grid reaches $0.9$); the $a>1$ observer scan above
diagnoses the nominal observer and is not used as a neural extrapolation claim.
The operating condition changes at $t=1$ s from $(a,d)=(0.5,5)$ to
$a\in\{0.1,0.2,0.4,0.6,0.9\}$ and
$d\in\{0,2,4,6,8\}$ through the original rate limiters.  Across 25 cases, the
expert, student, and tuned observer achieve mean one-second post-transition
RMSE values 0.00751, 0.00889, and 0.02888.  The student has no task failure and
outperforms the observer in 22 cases; the observer fails at $(0.1,6)$ and
$(0.1,8)$.

\begin{figure}[t]
\centering
\includegraphics[width=0.91\textwidth]{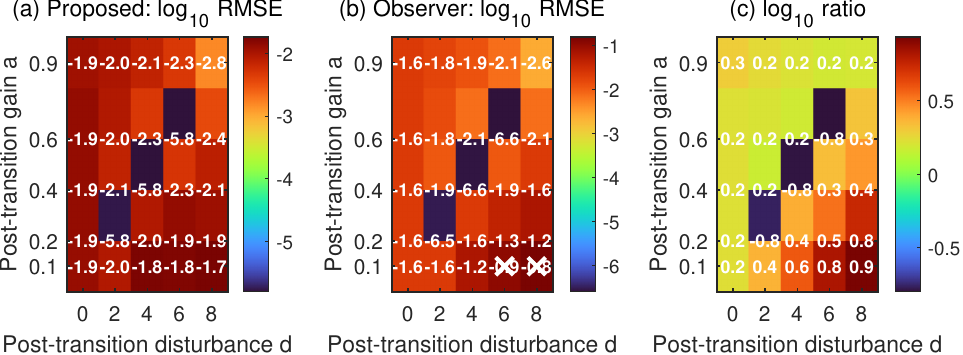}
\caption{Tracking performance over simultaneous post-transition values of
$a$ and $d$ with frozen observer gain $L=19000$.  Crosses denote observer task
failures.}
\label{fig:grid}
\end{figure}

\begin{figure}[t]
\centering
\includegraphics[width=0.91\textwidth]{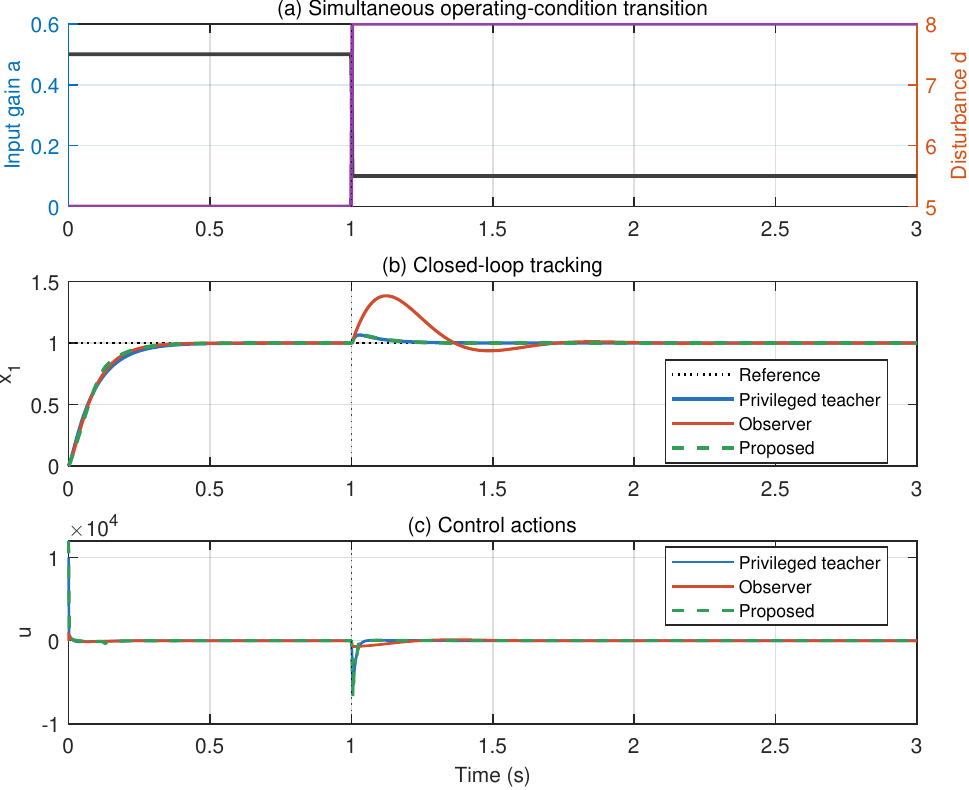}
\caption{Representative simultaneous multiplicative and additive uncertainty
transition.  The student does not receive $a$ or $d$.}
\label{fig:hard}
\end{figure}

At the hard transition $(0.5,5)\rightarrow(0.1,8)$, expert, student, and
observer RMSE values are 0.01745, 0.01821, and 0.15741, respectively.
Figure \ref{fig:hard} shows that the student reconstructs the expert-like
compensation from measurable causal history after the transient reveals the
new regime.

\subsection{Observer gain tradeoff}
The observer gain is scanned from $10^3$ to $10^5$ on four unseen seeds.  Mean
RMSE decreases from 0.04548 at $L=1000$ to 0.04153 at $L=19000$.  The
improvement from $L=15000$ to $19000$ is only $5.9\times10^{-5}$, whereas all
four runs diverge at the next tested gain $L=20000$ and at every higher gain.
Hence the selected observer lies at the observed edge of the performance--
robustness tradeoff rather than at a deliberately conservative setting.

\begin{figure}[t]
\centering
\includegraphics[width=0.86\textwidth]{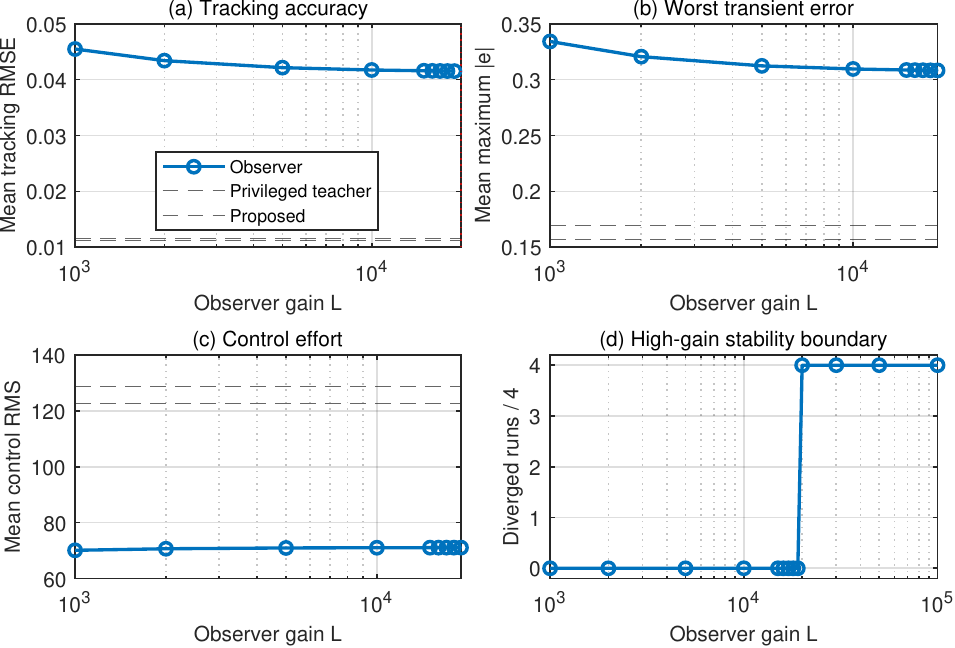}
\caption{Observer tracking, control effort, and tested high-gain stability
boundary.}
\label{fig:obs}
\end{figure}

\subsection{Unseen 60-s trials and robustness boundary}
Eight unseen seeds are simulated for 60 s.  The disturbance changes every 1 s
and the gain every 10 s.  No controller diverges.  Expert, student, and tuned
observer mean RMSE values are 0.00720, 0.00802, and 0.02587.  Thus the student
reduces RMSE by about 69\% relative to the tuned observer while staying close
to the expert.  Control RMS values are 80.30, 76.54, and 45.00; the improved
tracking requires expert-like action scale.

Sensor noise is applied symmetrically.  With
$\sigma_{x_2}=10\sigma_{x_1}$, the student has no task failure for
$\sigma_{x_1}\le10^{-4}$.  At $\sigma_{x_1}=10^{-3}$ it reaches saturation and
fails the difficult case.  The causal backward difference is the dominant
noise-sensitive component.  This result defines a present robustness boundary
and motivates a filtered derivative; it is not hidden as a tuning artifact.

\begin{figure}[t]
\centering
\includegraphics[width=0.84\textwidth]{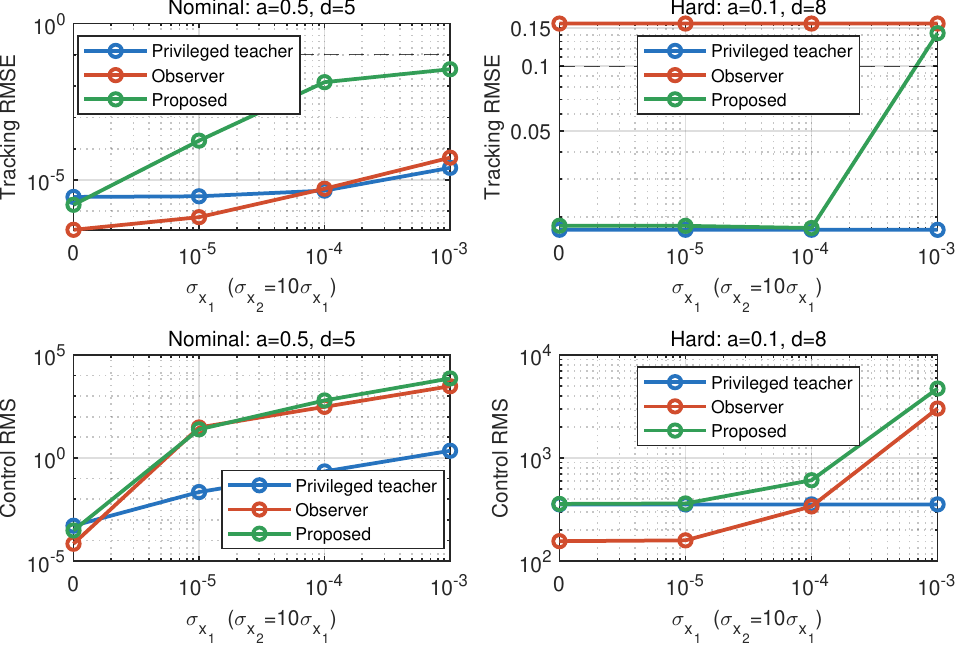}
\caption{Measurement-noise boundary of the structured student.}
\label{fig:noise}
\end{figure}

\subsection{What the experiments do and do not establish}
The experiments support H1--H5 within the declared second-order simulation
domain.  They do not prove extrapolation outside the sampled $a,d$ ranges,
robustness to arbitrary switching, or stability under sustained saturation.
Nor do they prove that every conventional observer or adaptive controller is
inferior.  Their role is to show that the proposed problem transformation is
effective, interpretable, and compatible with a nontrivial stability
certificate.

\section{Discussion, Open Questions, and Conclusion}

\subsection{Interpretation}
The final controller can legitimately be described as a conventional
mechanism-based action law plus a learned inverse-gain identifier.  This
interpretation strengthens rather than weakens the paper: it exposes exactly
what information the data supply.  The learner is not asked to rediscover
proportional feedback, infer two hidden quantities independently, and remain
stable all at once.  Privileged demonstrations define a desirable multi-regime
closed-loop distribution; physical algebra removes the matched disturbance;
history supports the remaining gain inference; and a certificate converts a
relative estimation range into a closed-loop statement.

The viewpoint also clarifies why privileged variables may be used to generate
samples but not fed to the deployed network.  Giving $a,d$ to the expert during
development is analogous to deliberately setting test conditions and tuning a
regime-aware controller.  Giving them to the student at deployment would
change the problem and merely reproduce that regime-aware controller.  In the
recommended training variant, even the latent label is recovered from expert
action and ordinary trajectories, although those trajectories still owe their
quality and diversity to privileged expert generation.

The expanded observer study sharpens the motivation without overstating it.
For the implemented nominal DOB, the unknown gain appears inside the lumped
disturbance as $(a-1)x_2$, so its estimation error is driven by the control
action.  Small $a$ and large $d$ expose a performance gap, while increasing
$L=19000$ to $20000$ changes every tested $a<1$ case from stable to divergent.
This does not establish that data-driven control is the only solution.  It
establishes that the missing regime adaptation is real and that one fixed
high-gain tuning is not a uniform remedy; privileged-data learning is the
remedy investigated here.

\subsection{Open research questions}
The second-order result exposes several questions that deserve separate work.
\begin{enumerate}
\item \emph{Higher-order task reduction:} which relative-degree, matching, or
normal-form conditions permit hidden disturbances to be eliminated while a
small identifiable latent remains?
\item \emph{Data design:} how should regimes and excitation be chosen to reduce
the history-space covering radius while controlling expert-generation cost?
\item \emph{On-policy correction:} can safe expert queries or aggregated
rollouts reduce student-induced distribution shift without requiring
privileged variables during ordinary operation?
\item \emph{Certificate-aware training:} can $\rho$ constraints, saturation,
noise filtering, and common-Lyapunov margins be included directly in the loss
or architecture?
\item \emph{Industrial validation:} can commissioning tests provide enough
privileged regime metadata to train a controller that improves an existing
product without compromising its baseline safety layer?
\end{enumerate}

\subsection{Limitations}
The common quadratic certificate is sufficient and conservative.  Its
regional relative-gain assumption is empirically frequent but not enforced for
all possible inputs.  Saturation and arbitrary parameter jumps are outside the
theorem.  Backward differentiation limits measurement-noise tolerance.  The
expert can also be suboptimal, and a student cannot be expected to exceed the
information and trajectory quality supplied during training without an
additional objective or exploration mechanism.

\subsection{Conclusion}
This paper presented a route from privileged multi-regime control to a
deployable controller for simultaneous unknown gain and additive disturbance.
The route combines realizability diagnosis, causal history, exact sampled-data
task reduction, action-derived supervision, and conditional closed-loop
analysis.  Direct black-box action imitation fails despite reasonable offline
fit, whereas a conventional small network succeeds after the learning target
is chosen by the control structure.  The broader contribution is therefore a
testable perspective: privileged control knowledge should be transferred only
after its deployable information content is diagnosed and its task-relevant,
physically composable component is isolated.

\appendix
\section{Derivation and Reproducibility Details}

\subsection{From continuous plant to Simulink timing}
On the interval $[(k-1)T_c,kT_c)$, zero-order-held $a_k,d_k$ and the previous
state value give
\begin{equation}
x_{1,k}-x_{1,k-1}
=\int_{(k-1)T_c}^{kT_c}[a_kx_2(t)+d_k]dt.
\end{equation}
The benchmark's discrete integrator/evaluation order implements the forward
Euler value $x_2(t)\mapsto x_{2,k-1}$ over this update, producing
\eqref{eq:x1d}.  The control update acts on $x_2$ through \eqref{eq:x2d}.
Consequently $\delta x_{1,k}$ corresponds to the interval ending at $k$, not
the interval beginning at $k$.  All label generation, training features, and
the stability model use this same convention.

\subsection{Action-derived label error}
Suppose the recorded action and derivative contain errors
$\tilde u_k$ and $\nu_k$.  The label computed from data is
\begin{equation}
\tilde\alpha_k^u=
\frac{(u_{p,k}+\tilde u_k)/k_2-\Delta x_{2,k}}
     {q_k-\nu_k}.
\end{equation}
Using $u_{p,k}/k_2-\Delta x_{2,k}=\alpha_kq_k$,
\begin{equation}
\tilde\alpha_k^u-\alpha_k
=\frac{\tilde u_k/k_2+\alpha_k\nu_k}{q_k-\nu_k}. \label{eq:labelnoise}
\end{equation}
If $|q_k|\ge q_{\min}$ and $|\nu_k|\le\bar\nu<q_{\min}$, then
\begin{equation}
|\tilde\alpha_k^u-\alpha_k|
\le\frac{|\tilde u_k|/k_2+\alpha_U\bar\nu}
{q_{\min}-\bar\nu}. \label{eq:labelbound}
\end{equation}
This explains both the $q_{\min}$ screen and the importance of derivative
filtering in noisy experiments.

\subsection{Finite coverage versus generalization}
Let $\mathcal M$ be a compact reachable set of normalized histories, and let
the ideal latent map $f_\star$ and estimator $f_\theta$ be Lipschitz with
constants $L_\star,L_\theta$.  Let the dataset covering radius be
\begin{equation}
h_{\mathcal D}=\sup_{z\in\mathcal M}\min_{z_i\in\mathcal D}\|z-z_i\|.
\end{equation}
If the maximum sample error is $\epsilon_{\mathcal D}$, then for the nearest
sample $z_i$,
\begin{align}
|f_\theta(z)-f_\star(z)|
&\le |f_\theta(z)-f_\theta(z_i)|
 +|f_\theta(z_i)-f_\star(z_i)|\nonumber\\
&\quad+|f_\star(z_i)-f_\star(z)|\\
&\le\epsilon_{\mathcal D}+(L_\theta+L_\star)h_{\mathcal D}. \label{eq:cover}
\end{align}
This conditional inequality formalizes why denser \emph{effective} history
coverage can improve a uniform error bound.  It does not convert an average
training loss into a global guarantee, nor does merely repeating highly
correlated equilibrium samples reduce $h_{\mathcal D}$.

Combining a bound $|\hat\alpha-\alpha|\le\epsilon_\alpha$ with
$a\in[a_L,a_U]$ gives
\begin{equation}
|\rho-1|=a|\hat\alpha-\alpha|\le a_U\epsilon_\alpha. \label{eq:rhobound}
\end{equation}
Therefore the certified interval is guaranteed if
$[1-a_U\epsilon_\alpha,1+a_U\epsilon_\alpha]\subseteq[\rho_L,\rho_U]$,
subject to positivity and the same timing assumptions.  This is the explicit
bridge from a uniform learned-latent error to Theorem 1.

\subsection{Numerical certificate audit}
The released source includes the full-precision matrix $P$, endpoint residuals,
and summary values.  A minimal audit performs:
\begin{enumerate}
\item verify $P=P^T$ and $\lambda_{\min}(P)>0$;
\item construct \eqref{eq:Ad} at $\rho=0.1$ and $4$;
\item verify the minimum eigenvalue of
$P-A_d(\rho)^TPA_d(\rho)$ is positive at both endpoints;
\item invoke the matrix-convexity lemma for all intermediate $\rho$.
\end{enumerate}
The certificate tests the implemented unsaturated recursion; it is independent
of any claim that the neural network globally approximates $1/a$.

\subsection{Recommended reporting checklist}
For future plants, report: the exact student information set; timing and delay
conventions; regime ranges and switching/rate limits; conflict diagnostics;
trajectory-level splits; latent-label conditioning threshold; saturation and
noise rules; observer tuning protocol; certificate interval; empirical fraction
inside that interval; and every failure.  These items make privileged
teacher--student control falsifiable and reproducible.

\small
\bibliographystyle{unsrt}
\bibliography{references}

@article{chen2016dobc,
  author={Wen-Hua Chen and Jun Yang and Lei Guo and Shihua Li},
  title={Disturbance-Observer-Based Control and Related Methods---An Overview},
  journal={IEEE Transactions on Industrial Electronics},
  year={2016}, volume={63}, number={2}, pages={1083--1095},
  doi={10.1109/TIE.2015.2478397}}

@article{vapnik2009lupi,
  author={Vladimir Vapnik and Akshay Vashist},
  title={A New Learning Paradigm: Learning Using Privileged Information},
  journal={Neural Networks}, year={2009}, volume={22}, number={5--6},
  pages={544--557}, doi={10.1016/j.neunet.2009.06.042}}

@inproceedings{chen2020learning,
  author={Dian Chen and Brady Zhou and Vladlen Koltun and Philipp Kr{\"a}henb{\"u}hl},
  title={Learning by Cheating}, booktitle={Proceedings of the Conference on Robot Learning},
  series={Proceedings of Machine Learning Research}, volume={100}, pages={66--75},
  year={2020}, url={https://proceedings.mlr.press/v100/chen20a.html}}

@inproceedings{kumar2021rma,
  author={Ashish Kumar and Zipeng Fu and Deepak Pathak and Jitendra Malik},
  title={{RMA}: Rapid Motor Adaptation for Legged Robots},
  booktitle={Proceedings of Robotics: Science and Systems}, year={2021},
  doi={10.15607/RSS.2021.XVII.011}}

@inproceedings{fu2023deep,
  author={Zipeng Fu and Xuxin Cheng and Deepak Pathak},
  title={Deep Whole-Body Control: Learning a Unified Policy for Manipulation and Locomotion},
  booktitle={Proceedings of the 6th Conference on Robot Learning},
  series={Proceedings of Machine Learning Research}, volume={205}, pages={138--149},
  year={2023}, url={https://proceedings.mlr.press/v205/fu23a.html}}

@inproceedings{paluch2025anc,
  author={Marcin Paluch and Florian Bolli and Pehuen Moure and Xiang Deng and Tobi Delbruck},
  title={{A-NC}: Adaptive Neural Control with Implicit Online Inference of Privileged Parameters},
  booktitle={Proceedings of the 7th Annual Learning for Dynamics and Control Conference},
  series={Proceedings of Machine Learning Research}, volume={283}, pages={987--998},
  year={2025}, url={https://proceedings.mlr.press/v283/paluch25a.html}}

@inproceedings{shenfeld2023tgrl,
  author={Idan Shenfeld and Zhang-Wei Hong and Aviv Tamar and Pulkit Agrawal},
  title={{TGRL}: An Algorithm for Teacher Guided Reinforcement Learning},
  booktitle={Proceedings of the 40th International Conference on Machine Learning},
  series={Proceedings of Machine Learning Research}, volume={202}, pages={31077--31093},
  year={2023}, url={https://proceedings.mlr.press/v202/shenfeld23a.html}}

@article{cai2024provable,
  author={Yang Cai and Xiangyu Liu and Argyris Oikonomou and Kaiqing Zhang},
  title={Provable Partially Observable Reinforcement Learning with Privileged Information},
  journal={arXiv preprint arXiv:2412.00985}, year={2024},
  url={https://arxiv.org/abs/2412.00985}}

@article{kim2025realizable,
  author={Yujin Kim and Nathaniel Chin and Arnav Vasudev and Sanjiban Choudhury},
  title={Distilling Realizable Students from Unrealizable Teachers},
  journal={arXiv preprint arXiv:2505.09546}, year={2025},
  url={https://arxiv.org/abs/2505.09546}}

@inproceedings{ross2011dagger,
  author={St{\'e}phane Ross and Geoffrey Gordon and Drew Bagnell},
  title={A Reduction of Imitation Learning and Structured Prediction to No-Regret Online Learning},
  booktitle={Proceedings of the Fourteenth International Conference on Artificial Intelligence and Statistics},
  series={Proceedings of Machine Learning Research}, volume={15}, pages={627--635},
  year={2011}, url={https://proceedings.mlr.press/v15/ross11a.html}}

@article{rusu2015policy,
  author={Andrei A. Rusu and Sergio Gomez Colmenarejo and Caglar Gulcehre and Guillaume Desjardins and James Kirkpatrick and Razvan Pascanu and Volodymyr Mnih and Koray Kavukcuoglu and Raia Hadsell},
  title={Policy Distillation}, journal={arXiv preprint arXiv:1511.06295},
  year={2015}, url={https://arxiv.org/abs/1511.06295}}

@article{hou2013mfac,
  author={Zhongsheng Hou and Shangtai Jin},
  title={Data-Driven Model-Free Adaptive Control for a Class of {MIMO} Nonlinear Discrete-Time Systems},
  journal={IEEE Transactions on Neural Networks and Learning Systems}, year={2013},
  volume={24}, number={6}, pages={887--901}, doi={10.1109/TNNLS.2013.2245671}}

@article{depersis2020formulas,
  author={Claudio De Persis and Pietro Tesi},
  title={Formulas for Data-Driven Control: Stabilization, Optimality, and Robustness},
  journal={IEEE Transactions on Automatic Control}, year={2020}, volume={65},
  number={3}, pages={909--924}, doi={10.1109/TAC.2019.2901658}}

@article{dawson2023survey,
  author={Charles Dawson and Sicun Gao and Chuchu Fan},
  title={Safe Control With Learned Certificates: A Survey of Neural Lyapunov, Barrier, and Contraction Methods},
  journal={IEEE Transactions on Robotics}, year={2023}, volume={39}, number={3},
  pages={1749--1767}, doi={10.1109/TRO.2022.3232542}}

@article{osa2018imitation,
  author={Takayuki Osa and Joni Pajarinen and Gerhard Neumann and J. Andrew Bagnell and Pieter Abbeel and Jan Peters},
  title={An Algorithmic Perspective on Imitation Learning},
  journal={Foundations and Trends in Robotics}, year={2018}, volume={7},
  number={1--2}, pages={1--179}, doi={10.1561/2300000053}}

@article{narendra1990nn,
  author={Kumpati S. Narendra and Kannan Parthasarathy},
  title={Identification and Control of Dynamical Systems Using Neural Networks},
  journal={IEEE Transactions on Neural Networks}, year={1990}, volume={1},
  number={1}, pages={4--27}, doi={10.1109/72.80202}}

@article{karniadakis2021piml,
  author={George Em Karniadakis and Ioannis G. Kevrekidis and Lu Lu and Paris Perdikaris and Sifan Wang and Liu Yang},
  title={Physics-Informed Machine Learning}, journal={Nature Reviews Physics},
  year={2021}, volume={3}, pages={422--440}, doi={10.1038/s42254-021-00314-5}}

@article{brunke2022safe,
  author={Lukas Brunke and Melissa Greeff and Adam W. Hall and Zhaocong Yuan and Siqi Zhou and Jacopo Panerati and Angela P. Schoellig},
  title={Safe Learning in Robotics: From Learning-Based Control to Safe Reinforcement Learning},
  journal={Annual Review of Control, Robotics, and Autonomous Systems}, year={2022},
  volume={5}, pages={411--444}, doi={10.1146/annurev-control-042920-020211}}

@inproceedings{tobin2017domain,
  author={Josh Tobin and Rachel Fong and Alex Ray and Jonas Schneider and Wojciech Zaremba and Pieter Abbeel},
  title={Domain Randomization for Transferring Deep Neural Networks from Simulation to the Real World},
  booktitle={2017 IEEE/RSJ International Conference on Intelligent Robots and Systems},
  year={2017}, pages={23--30}, doi={10.1109/IROS.2017.8202133}}

@article{hwangbo2019skills,
  author={Jemin Hwangbo and Joonho Lee and Alexey Dosovitskiy and Dario Bellicoso and Vassilios Tsounis and Vladlen Koltun and Marco Hutter},
  title={Learning Agile and Dynamic Motor Skills for Legged Robots},
  journal={Science Robotics}, year={2019}, volume={4}, number={26},
  pages={eaau5872}, doi={10.1126/scirobotics.aau5872}}

@inproceedings{chen2025slr,
  author={Shiyi Chen and Zeyu Wan and Shiyang Yan and Chun Zhang and Weiyi Zhang and Qiang Li and Debing Zhang and Fasih Ud Din Farrukh},
  title={{SLR}: Learning Quadruped Locomotion without Privileged Information},
  booktitle={Proceedings of the 8th Conference on Robot Learning},
  series={Proceedings of Machine Learning Research}, volume={270},
  pages={3212--3224}, year={2025},
  url={https://proceedings.mlr.press/v270/chen25e.html}}

@article{disaro2025uio,
  author={Giorgia Disar{\`o} and Maria Elena Valcher},
  title={On the Equivalence of Model-Based and Data-Driven Approaches to the Design of Unknown-Input Observers},
  journal={IEEE Transactions on Automatic Control}, year={2025}, volume={70},
  number={3}, pages={2074--2081}, doi={10.1109/TAC.2024.3482928}}

@inproceedings{nadali2025snsr,
  author={Alireza Nadali and Ashutosh Trivedi and Majid Zamani},
  title={Stochastic Neural Simulation Relations for Control Transfer},
  booktitle={Proceedings of the International Conference on Neuro-symbolic Systems},
  series={Proceedings of Machine Learning Research}, volume={288},
  pages={597--620}, year={2025},
  url={https://proceedings.mlr.press/v288/nadali25a.html}}
\end{document}